\documentclass[a4paper]{jpconf}
\usepackage[dvipdfmx]{graphicx}
\usepackage{color}
\begin{document}
\title{Performance of VUV-sensitive MPPC for Liquid Argon Scintillation Light}
\author{
T.Igarashi,
S.Naka,
M.Tanaka,
T.Washimi,
K.Yorita
}

\address{Waseda University, Tokyo, Japan}
\ead{masashi.tanaka@aoni.waseda.jp, kohei.yorita@waseda.jp}

\begin{center}
\begin{abstract}
 A new type of the Multi-Pixel Photon Counter (MPPC), sensitive to Vacuum Ultra-Violet (VUV)
light (wavelength $\lambda$ $<$ 150 nm), is recently developed and produced by Hamamatsu
Photonics K.K..
The basic properties of the new MPPC are measured at cryogenic facility of Waseda university
using liquid nitrogen. Temperature dependence of breakdown voltage, capacitance, and dark count
rate of the MPPC are also evaluated. 
In addition, the absolute photon detection efficiency (PDE) for liquid argon (LAr) scintillation 
light ($\lambda$ = 128 nm) for the latest model of MPPC is estimated to be 12\% by using $^{241}$Am $\alpha$-ray source. 
Based on these basic measurements a possible application of the new MPPC to LAr detector for 
dark matter search is discussed.
\end{abstract}
\end{center}


\section{Introduction}
Liquid argon (LAr) is known to be an excellent target material for various particle physics
experiments, such as neutrino physics, nucleon decay, and direct search for weakly interacting
massive particle (WIMP dark matter). For WIMP search, use of its ionization and scintillation
signals in addition to scintillation pulse shape discrimination provides a strong rejection power
between the electron-recoil (main background) and the nuclear-recoil (WIMP signal) events.
On the other hand the scintillation light of LAr, particularly for the purpose of particle physics
 experiment, is difficult to detect with normal photo-sensors since it is required for such
photo-sensors to have the abilities of:
\begin{itemize}
\item detecting 128 nm VUV photon of LAr scintillation light;
\item operational under LAr temperature at $-$186 $^\circ$C;
\item operational with relatively low voltage to avoid discharge in pure argon;
\item counting one photo-electron for tiny physics signals and powerful background rejection.
\end{itemize}
Due to these technical requirements, the conventional procedure to detect LAr scintillation light
so far is to convert the 128 nm light to visible light (420 nm) using tetraphenyl butadiene (TPB)
as a wavelength shifter and those converted photons are detected by cryogenic photo-multiplier
tubes (PMT).\\

As a part of the ANKOK project~\cite{Tanaka:2013oma}, a direct dark matter search experiment using
double phase Ar detector technique, we are under development of a new procedure to detect the LAr
scintillation light using Multi-Pixel Photon Counters (MPPCs). Normal MPPCs have high efficiency
peaking at 400 to 500 nm and in general no sensitivity for VUV light (here below 150 nm). For
liquid xenon (LXe) scintillation light (wavelength of 175 nm), improved MPPCs have been developed
and are already in practical stage used by MEG experiment~\cite{MEG}. A new MPPC, further extended
the sensitivity to below 150 nm, is currently under development by
Hamamatsu Photonics K.K.~\cite{hamamatsu0} and thus this kind of new MPPC may be considered to open new
style of liquid or gaseous argon scintillation detector in the near future.\\
\indent In this paper, basic properties of the newly improved MPPC, its performance for detecting
LAr scintillation light, and its physics application are reported and discussed.

\section{The VUV-sensitive MPPC}
The new MPPCs, type number "3X3MM-50UM VUV2", "3X3MM-50UM VUV3" and "3X3MM-100UM VUV3"~\cite{hamamatsu1}, 
were developed for the direct detection of VUV light below 150 nm by Hamamatsu Photonics K.K. and tested 
at Waseda university campus under cryogenic environment.
"VUV2" type was produced in April 2014 and their basic parameters are similar to commercially available MPPC, 
type number "S12572-33-050C"~\cite{hamamatsu2}.
"VUV3" type, produced in April 2015, is cross-talk suppressed model similar to type number "S13360-3050CS"~\cite{hamamatsu3}.
The chip size is 3 mm $\times$ 3 mm, and the labels of  "-50UM" and "-100UM" denote pixel size of 50 $\mu$m and 100 $\mu$m, respectively. 

Basic parameters of these MPPCs measured by Hamamatsu Photonics K.K. are summarized in Table~\ref{table1}. 
These measurements are all done at room temperature of 25$^\circ$C.

\begin{table}[h]
\caption{\label{table1}List of the new VUV-sensitive MPPCs and their basic properties measured
by Hamamatsu Photonics at room temperature of 25$^\circ$C.}
\begin{center}
\begin{tabular}{c|c|c|c|c|c}
MPPC        & Type No.         & Serial No.  & Bias voltage & Gain                & Dark counts \\
\hline
V2-50UM-(1) & 3X3MM-50UM VUV2  & A0010       & 66.65 V      & 1.25 $\times 10^{6}$ & 572 kHz \\
V2-50UM-(2) & 3X3MM-50UM VUV2  & A0011       & 66.77 V      & 1.25 $\times 10^{6}$ & 701 kHz \\
V3-50UM     & 3X3MM-50UM VUV3  & A0011       & 54.87 V      & 2.00 $\times 10^{6}$ & 674 kHz \\
V3-100UM    & 3X3MM-100UM VUV3 & A0003       & 53.78 V      & 5.50 $\times 10^{6}$ & 553 kHz \\
\end{tabular}
\end{center}
\end{table}

\section{Basic Property of VUV-sensitive MPPC}
\begin{figure}[h]
\begin{center}
\begin{minipage}{6cm}
\includegraphics[width=6cm]{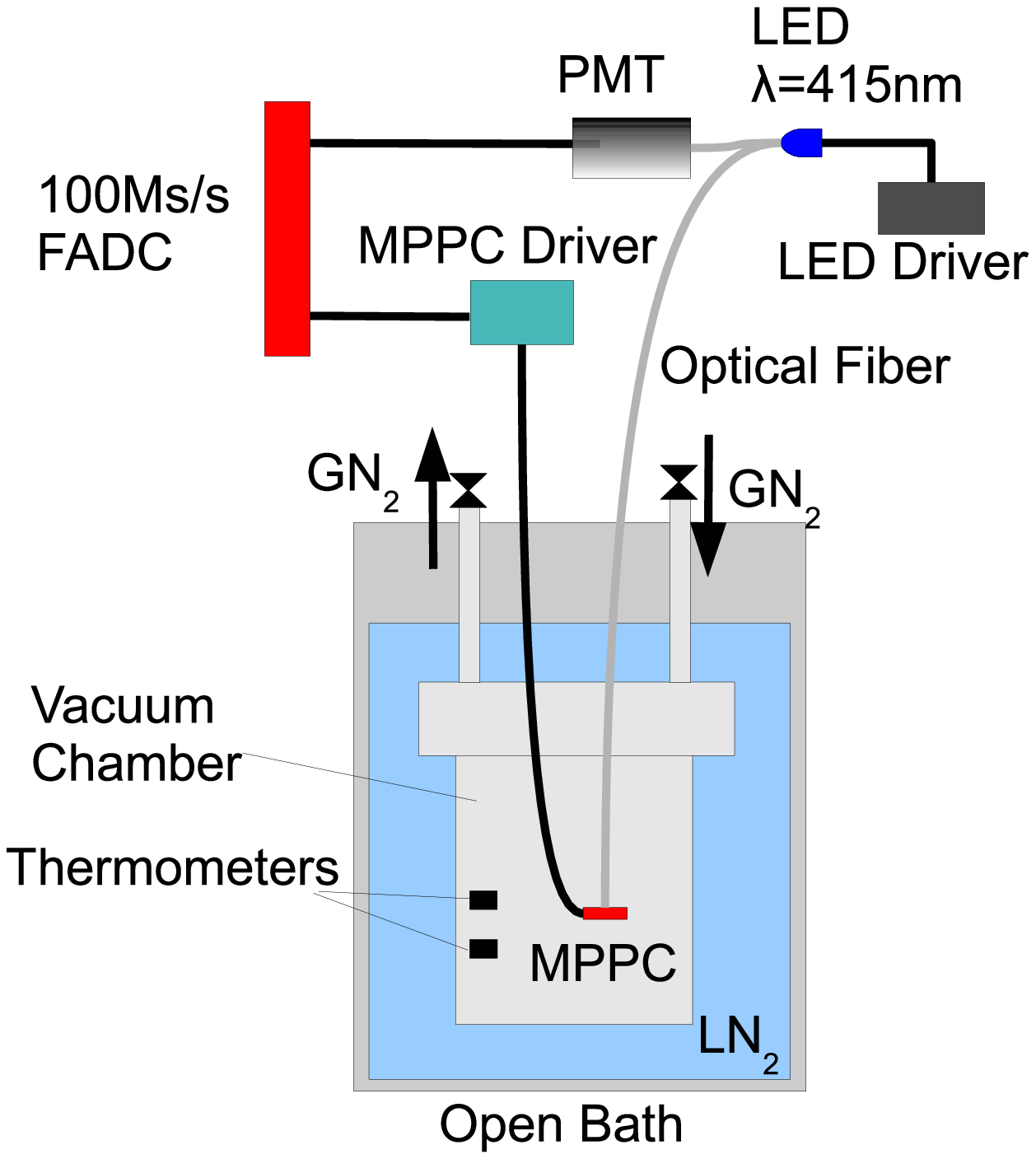}
\end{minipage}
\begin{minipage}{8cm}
\includegraphics[width=9cm]{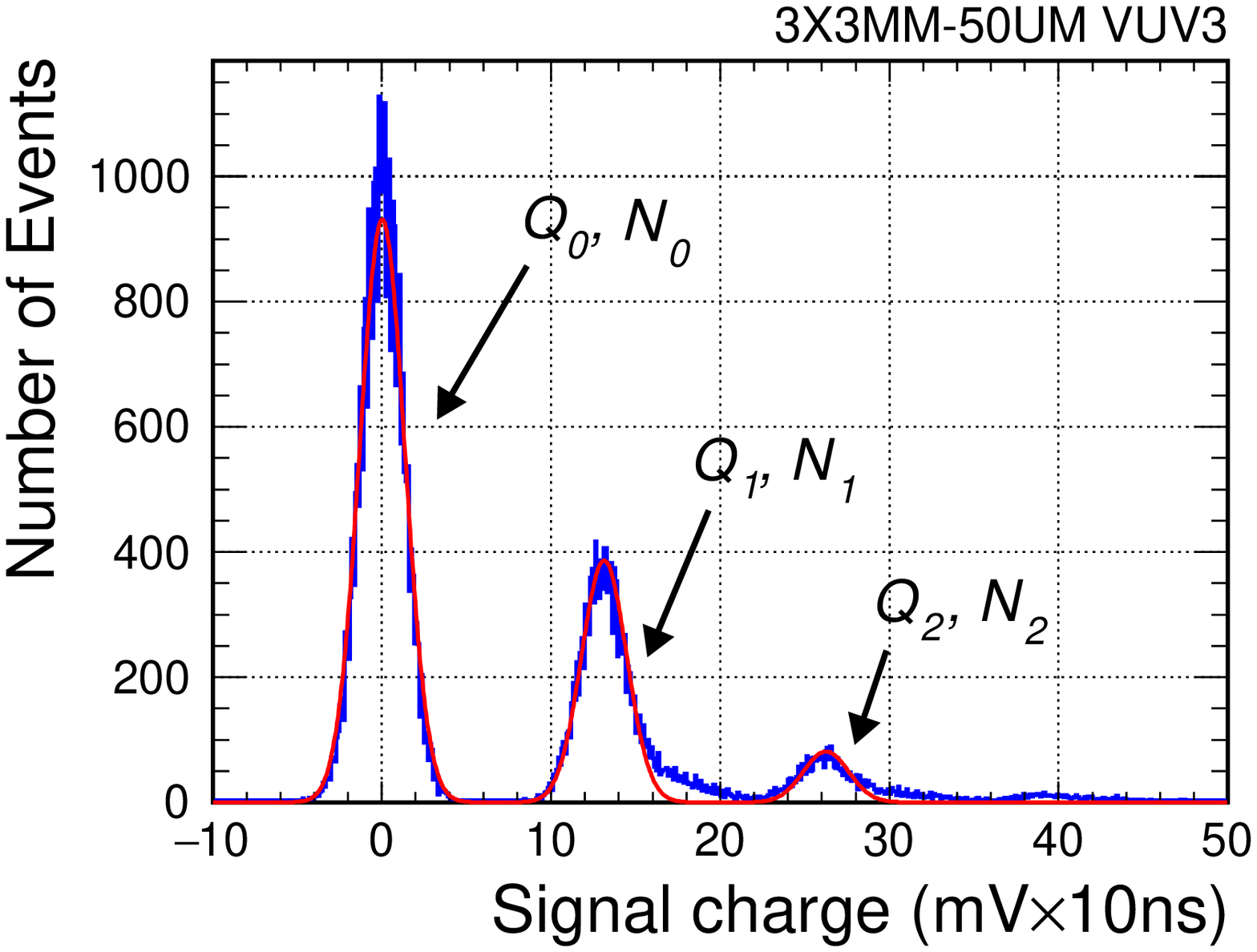}
\end{minipage}
\end{center}
\caption{\label{Prop1}Schematic diagram of the setup for MPPC basic property measurement
at cryogenic temperature (left) and MPPC (3X3MM-50UM VUV3) signal charge distribution obtained by the 415 nm
LED light at $-$190 $^\circ$C (right).}
\end{figure}

\begin{figure}[h]
\begin{center}
\includegraphics[width=17cm]{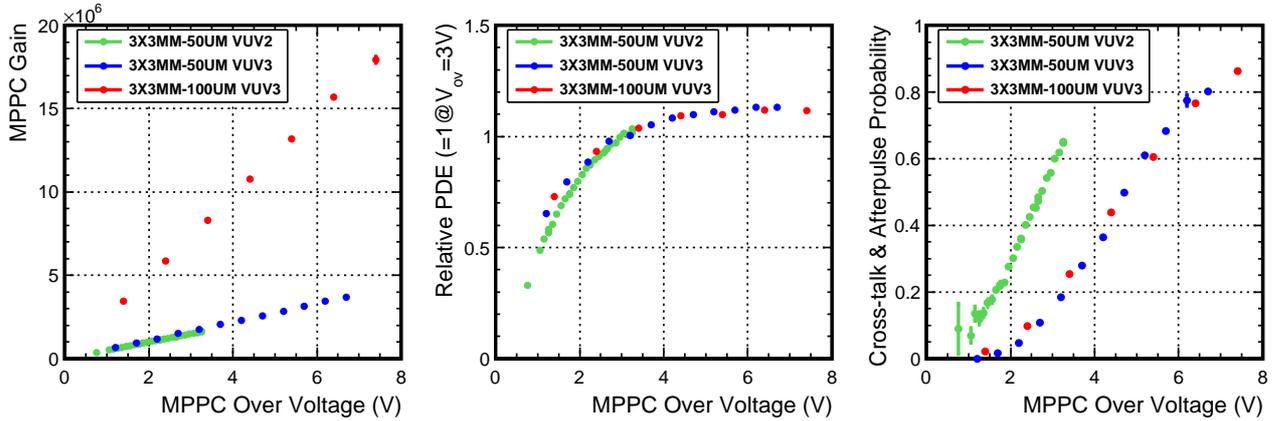}
\end{center}
\caption{\label{Prop2} MPPC gain (left), relative PDE (middle), and cross-talk and afterpulse probability (right)
as a function of the MPPC over voltage at $-$190 $^\circ$C measured by the 415 nm LED light.}
\end{figure}

In addition to the properties in Table~\ref{table1} measured at room temperature,
we have extensively measured the performance of 3 MPPCs (V2-50UM-(1), V3-50UM, and V3-100UM) at cryogenic temperature.
Left picture in Fig.~\ref{Prop1} shows schematic diagram of the test setup built
at Waseda University. The MPPC is mounted inside the vacuum
chamber filled with 1 bar of pure gas nitrogen and refrigerated by immersing the
chamber into the open bath filled with liquid nitrogen. Two thermometers located
close to the MPPC are used to monitor temperature around MPPC and the difference
of the two readout values ($\pm$5 $^\circ$C) is considered as uncertainty of the MPPC
temperature. An LED with wavelength of 415 nm is located outside of the chamber
at room temperature, and photons from the LED are injected to the surface of the MPPC
through the optical fiber and feed-through. The average light yield of the LED signal is
adjusted to $\sim$1 photon per pulse at the surface of MPPC. The LED pulse time width
is few ns which is much shorter than the MPPC time response ($\sim$40 ns for "-50UM").
The LED light yield per pulse is monitored by PMT (Hamamatsu H1161) and its stability
is found to be within 0.5\% throughout the measurement period. MPPC driver kit
(Hamamatsu C12332) is used to supply bias voltage and signal amplification (gain $G_{\rm amp} =10.9 \pm 0.1$).
The signal is digitized by CAEN FADC (V1724, 100 Ms/s) and the digitized data is acquired
and stored on PC. The FADC waveform is analyzed to extract the signal charge by integrating
the waveform in the time range of [$-$20 ns, 120 ns] (for  "-50UM") and [$-$20 ns, 500 ns] 
(for  "-100UM") from the LED pulse timing.\\
Right plot in Fig.~\ref{Prop1} shows an example of signal charge distribution in unit of
mV $\times$ 10 ns for 50000 events of data, obtained at $-$190 $^\circ$C with MPPC (V3-50UM) bias voltage
of 44.0 V. Each peak in the plot corresponds to 0, 1, and 2 photo-electrons. The signal charge
distribution is simultaneously fitted by three Gaussian distributions to estimate each mean charge
($Q_0$, $Q_1$, $Q_2$) and the number of events ($N_0$, $N_1$, $N_2$)
as described in Fig.~\ref{Prop1}. 
MPPC gain ($G$) is obtained from the difference between the peak charge
($Q_G\equiv Q_1-Q_0= Q_2-Q_1$) by using the relation of $Q_G = \e\times G\times G_{\rm amp}$ where e is
the elementary charge and the factor $G_{\rm amp}$ is the signal amplification gain.\\\\
As well known, the gain of the MPPC changes linearly as a function of over voltage ($V_{\rm ov}$)
even at $-$190 $^\circ$C as shown in left plot in Fig.~\ref{Prop2}. Note that for this measurement
the breakdown voltage ($V_{\rm bd}$) is obtained to be 41.31 V from the $V_{\rm bias}-G$ plot by
linearly extrapolating to the zero-gain point and already subtracted in Figs.~\ref{Prop2}. 
We also perform the same procedure for V2-50UM-(1) and V3-100UM, and these results are shown in Fig.~\ref{Prop2}.
\\
The number of photons injected from LED to the MPPC is assumed to follow Poisson distribution
with average $\mu_{\rm in}$. Thus the number of events with 0 photo-electron is described as
\begin{equation}
N_0 = N_{\rm all}\times e^{-\mu} = N_{\rm all}\times e^{-\mu_{\rm in}\times {\rm PDE}} ,
\end{equation}
where $N_{\rm all}$ is the total number of events and PDE is the photon detection efficiency.
In this setup absolute value of PDE can not be extracted because original $\mu_{\rm in}$ is unknown.
However the relative dependence on $V_{\rm ov}$ and temperature can be estimated.
Middle plot in Fig.~\ref{Prop2} shows relative PDE as a function of $V_{\rm ov}$, normalized
to unity at $V_{\rm ov}=$ 3.0 V.
 Because of the effect of 
the cross-talk and afterpulse which possibly occurs in the time window, the number of events observed in 1 photo-electron peak is reduced 
from Poisson expectation by the cross-talk and afterpulse probability ($X$) defined as follows,
\begin{equation}
N_1 = N_{\rm all}\times \mu e^{-\mu} \times (1-X).
\end{equation}
Cross-talk and afterpulse probability is evaluated from wider time window, integral range of [$-$20 ns, 600 ns], to collect afterpulses. 
As shown in right plot in Fig.~\ref{Prop2}, the cross-talk and afterpulse probability $X$ becomes higher
as imposed over voltage ($V_{\rm ov}$) increases, and becomes lower in VUV3 compared to VUV2 type at the same over voltage. 
\\

\begin{figure}[h]
\begin{center}
\begin{minipage}{8cm}
\includegraphics[width=8cm]{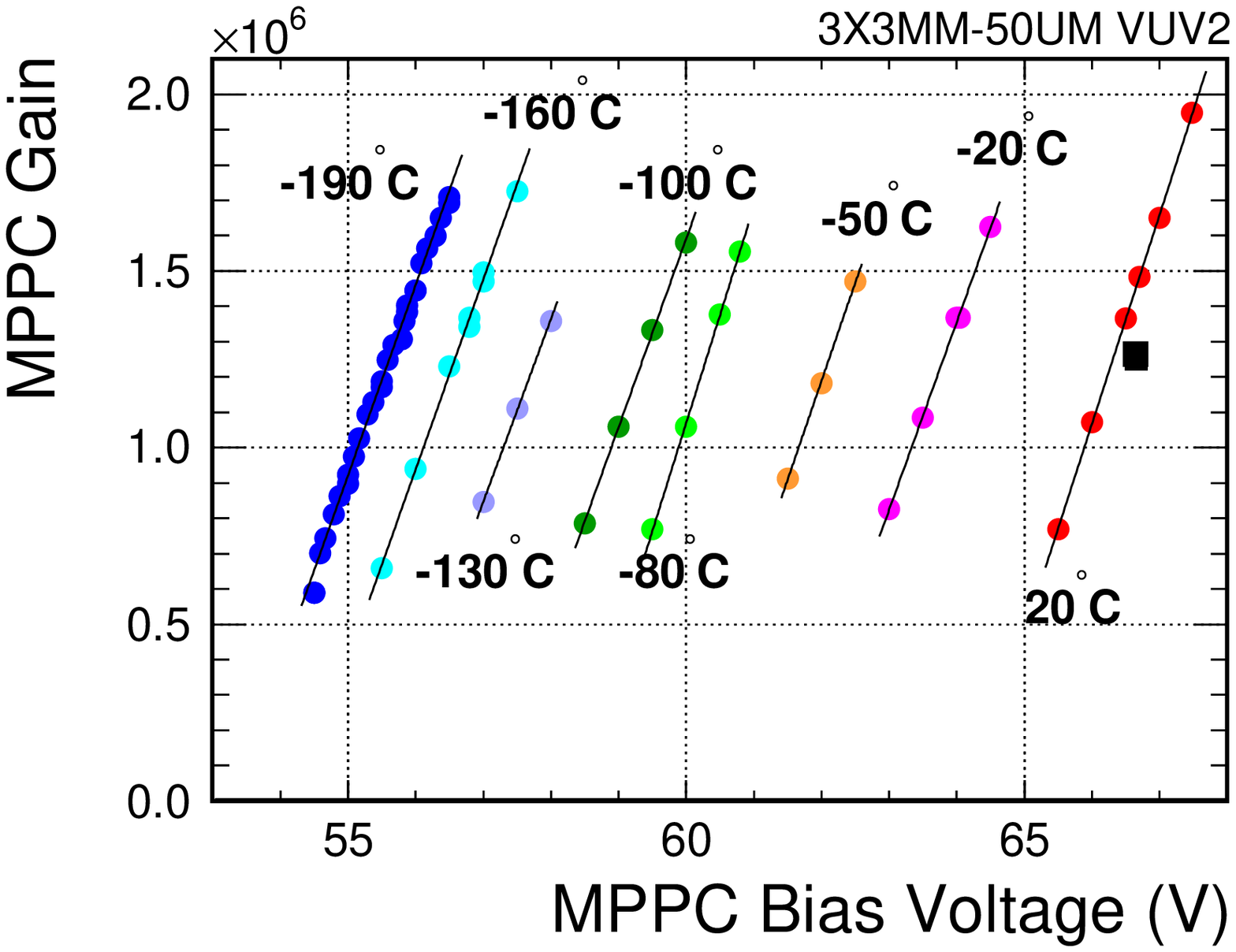}
\end{minipage}
\begin{minipage}{7.5cm}
\includegraphics[width=7.5cm]{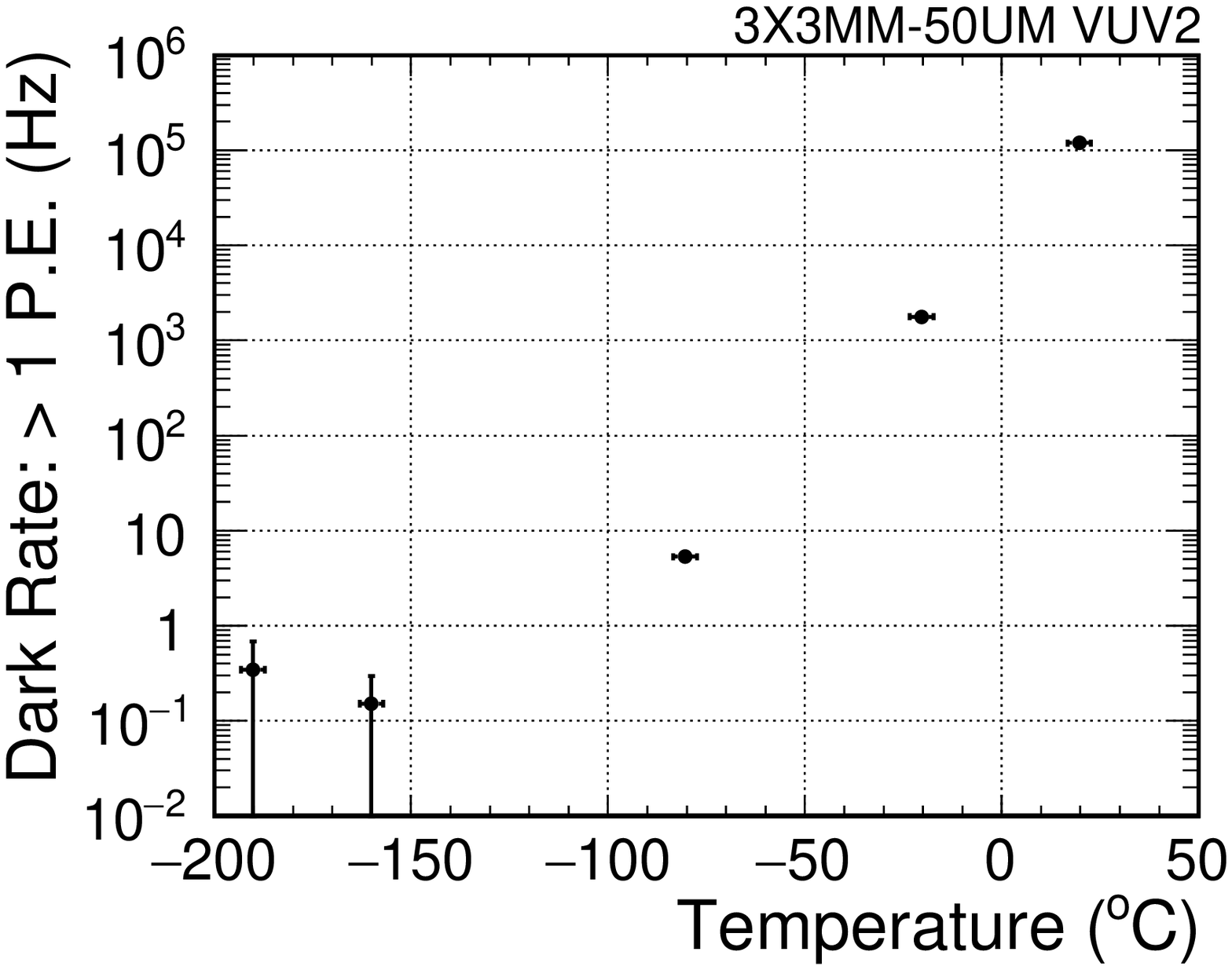}
\end{minipage}
\caption{\label{Prop3} 3X3MM-50UM VUV2 (A0010) gain as a function of over voltage in various operation temperature
(left). The closed square in the plot corresponds to the measured value by Hamamatsu Photonics K.K.
at 25 $^{\circ}$C. Right plot is dark count rate with threshold of 1 photo-electron at different
temperatures.}
\end{center}
\end{figure}

In order to evaluate wider range of temperature dependence of the new MPPC, we have further
measured gain and dark count rate at 8 different temperature points ($-190$, $-160$, $-130$,
$-100$, $-80$, $-50$, $-20$, $20$ $^{\circ}$C). Left plot in Fig.~\ref{Prop3} shows gain as
a function of bias voltage at each different temperature for V2-50UM-(1). The gain slope, which is proportional
to MPPC capacitance, has no dependence over temperature and found to be $\sim$ 85 fF 
while the breakdown voltage decreases along with temperature ($\sim$ 50 mV/$^{\circ}$C).
Right plot in Fig.~\ref{Prop3} shows the dark rate of V2-50UM-(1) counted with 1 photo-electron equivalent threshold
without the LED light signals. The rate is significantly reduced in lower temperature and becomes
less than 1 Hz at cryogenic temperature (below $\sim$150 $^{\circ}$C) where the upper bound is limited
by experimental conditions ($e.g.$ small light leak).

\section{Measurement of Liquid Argon Scintillation Light} \label{sec:LAr}
\begin{figure}[h]
\begin{center}
\includegraphics[width=10cm]{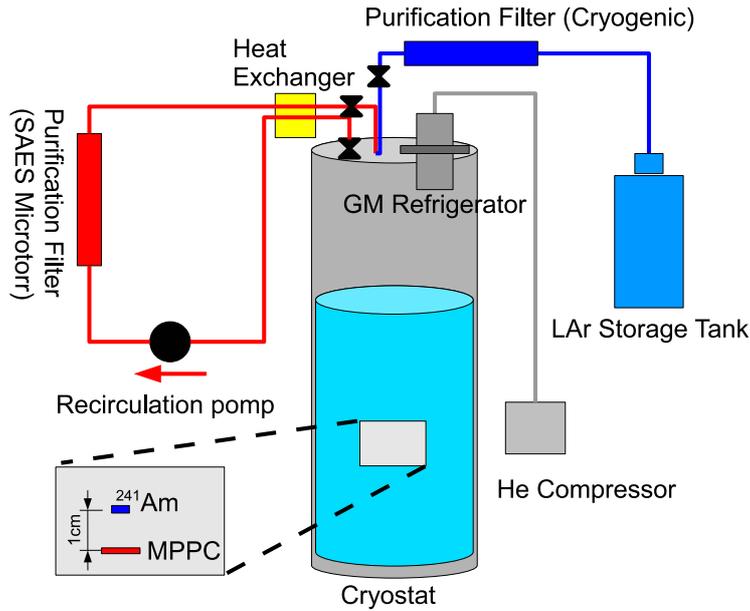}
\end{center}
\caption{\label{Liq1}Experimental apparatus for liquid argon test setup at Waseda university.}
\end{figure}

Figure~\ref{Liq1} shows schematic diagram of the setup for measuring LAr scintillation light.
Detail of the setup to obtain high purity LAr is described elsewhere \cite{Tanaka:2013oma}.
3 MPPCs (V2-50UM-(2), V3-50UM and V3-100UM) are 
immersed into the LAr and $^{241}$Am $\alpha$-ray source ($\sim$40 Bq) is placed at 1 cm apart from the MPPC.
MPPC power supply and signal readout for the setup are the same as in Fig.~\ref{Prop1}.

\begin{figure}[h]
\begin{center}
\includegraphics[width=17cm]{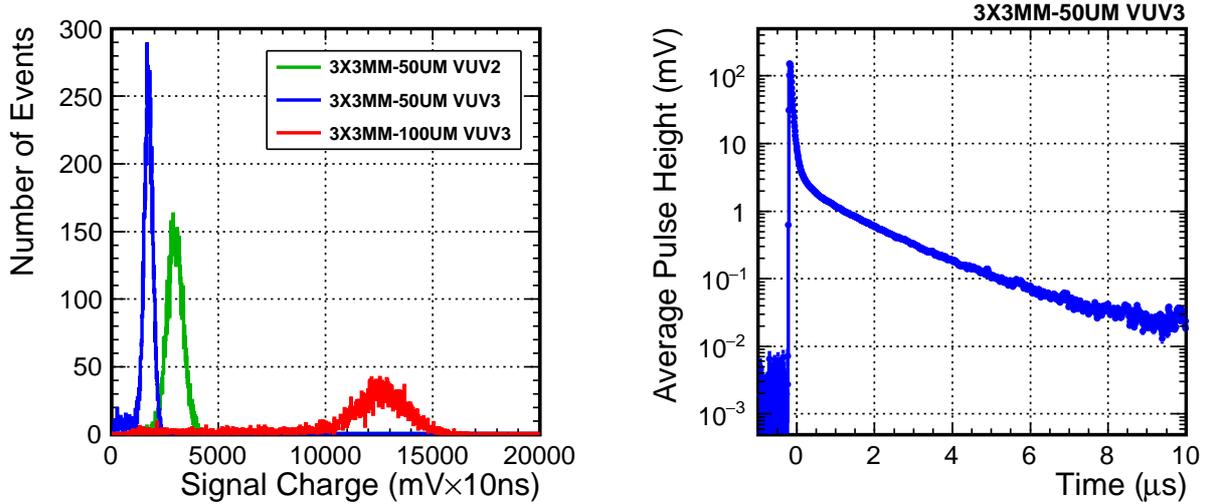}
\end{center}
\caption{\label{Liq2} Signal charge distributions for 3 different types of MPPCs (left) 
and average waveform (right) for one MPPC (V3-50UM) of argon
scintillation light by $^{241}$Am $\alpha$-ray (5.5 MeV) at $V_{\rm ov}\sim$ 3 V.}
\end{figure}

Left plot in Fig.~\ref{Liq2} shows signal charge distributions of 3 MPPCs at the same over voltage 
($V_{\rm ov}\sim$3 V) within [-20 ns, 10 $\mu$s] from
the trigger timing. A clear peak corresponds to 5.5 MeV $\alpha$-ray from $^{241}$Am and its 
signal rate is consistent with the expectation. The distribution is fitted into
single Gaussian function to obtain mean charge ($Q_{\alpha}$) and the energy resolution ($\sigma$/$Q_{\alpha}$). 
The results are summarized in Table~\ref{table2}. 
Improvement of energy resolution by cross-talk suppression (from VUV2 to VUV3) and change of pixel size 
(from 50 $\mu$m to 100 $\mu$m) are observed.
Right plot in Fig.~\ref{Liq2} shows average waveform for the signal events within 2 $\sigma$ around
the peak of the charge distribution. Scintillation light of LAr is known to have two components
with different decay time, fast ($\tau\sim$10 ns) and slow ($\tau\sim$1.5 $\mu$s) and the slow
fraction to the total charge for $\alpha$-ray is expected to be 0.2$\sim$0.3~\cite{Doke:1999ku}.
The waveform obtained by this measurement does have the two components, and their decay times
and the fraction are about consistent with expectations.\\

\begin{table}[h]
\caption{\label{table2} Mean charge and energy resolution of Fig.5 (left).}
\begin{center}
\begin{tabular}{c|c|c|c|c}
Type No.         & Serial No. & Bias voltage & Mean charge      & Energy resolution \\
                 &            & [V]          & [mV$\times$10ns] & [\%] \\
\hline
3X3MM-50UM VUV2  & A0011      &  56.20       & $2963\pm 5$      & $12.6 \pm 0.2$  \\
3X3MM-50UM VUV3  & A0011      &  44.85       & $1725\pm 3$      & $11.6 \pm 0.2$  \\
3X3MM-100UM VUV3 & A0003      &  45.27       & $12642\pm 17$    & $9.1 \pm 0.2$  \\
\end{tabular}
\end{center}
\end{table}


\begin{figure}[h]
\begin{center}
\begin{minipage}{7.5cm}
\includegraphics[width=8.2cm]{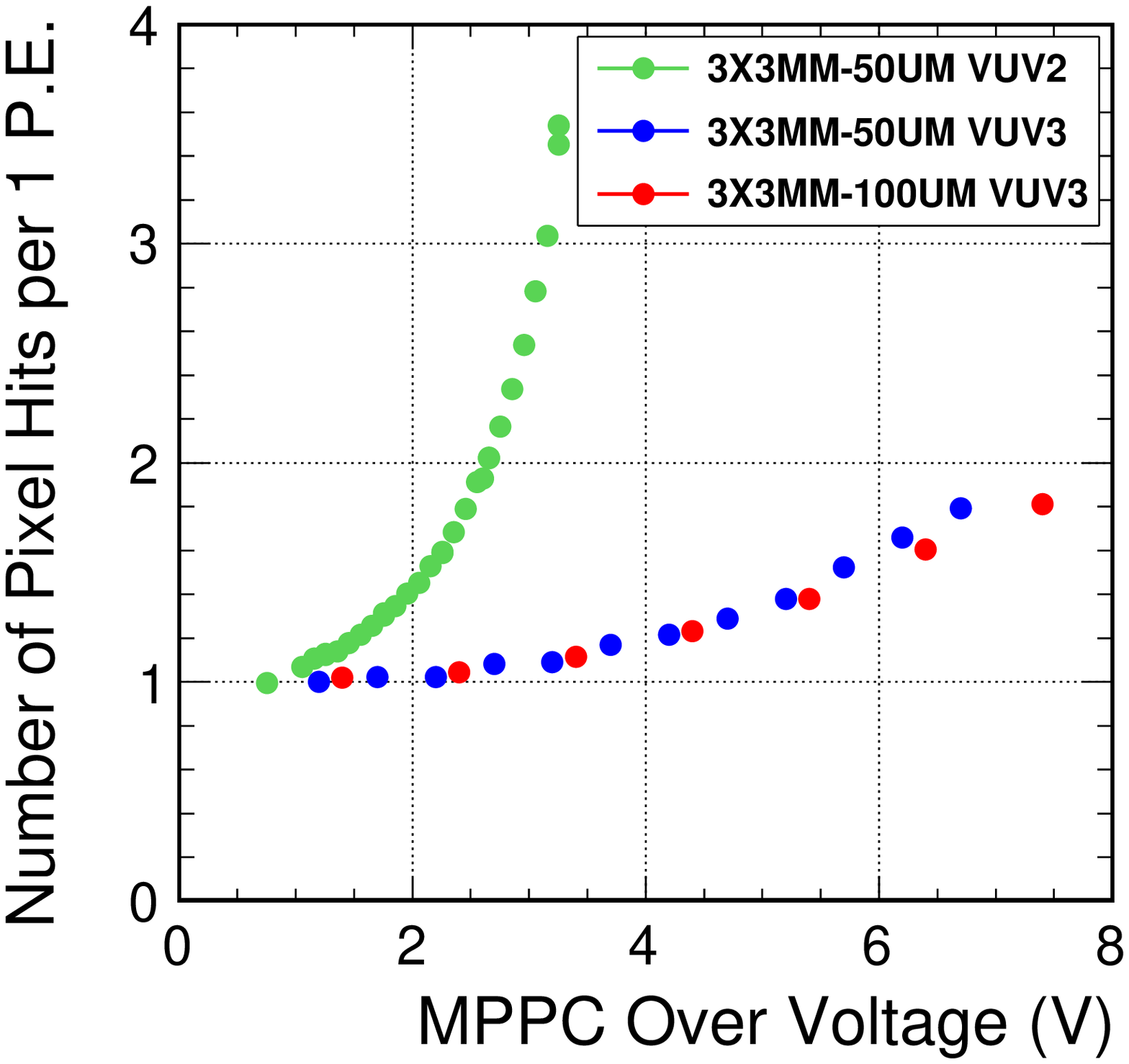}
\end{minipage}
\begin{minipage}{7.5cm}
\includegraphics[width=8.2cm]{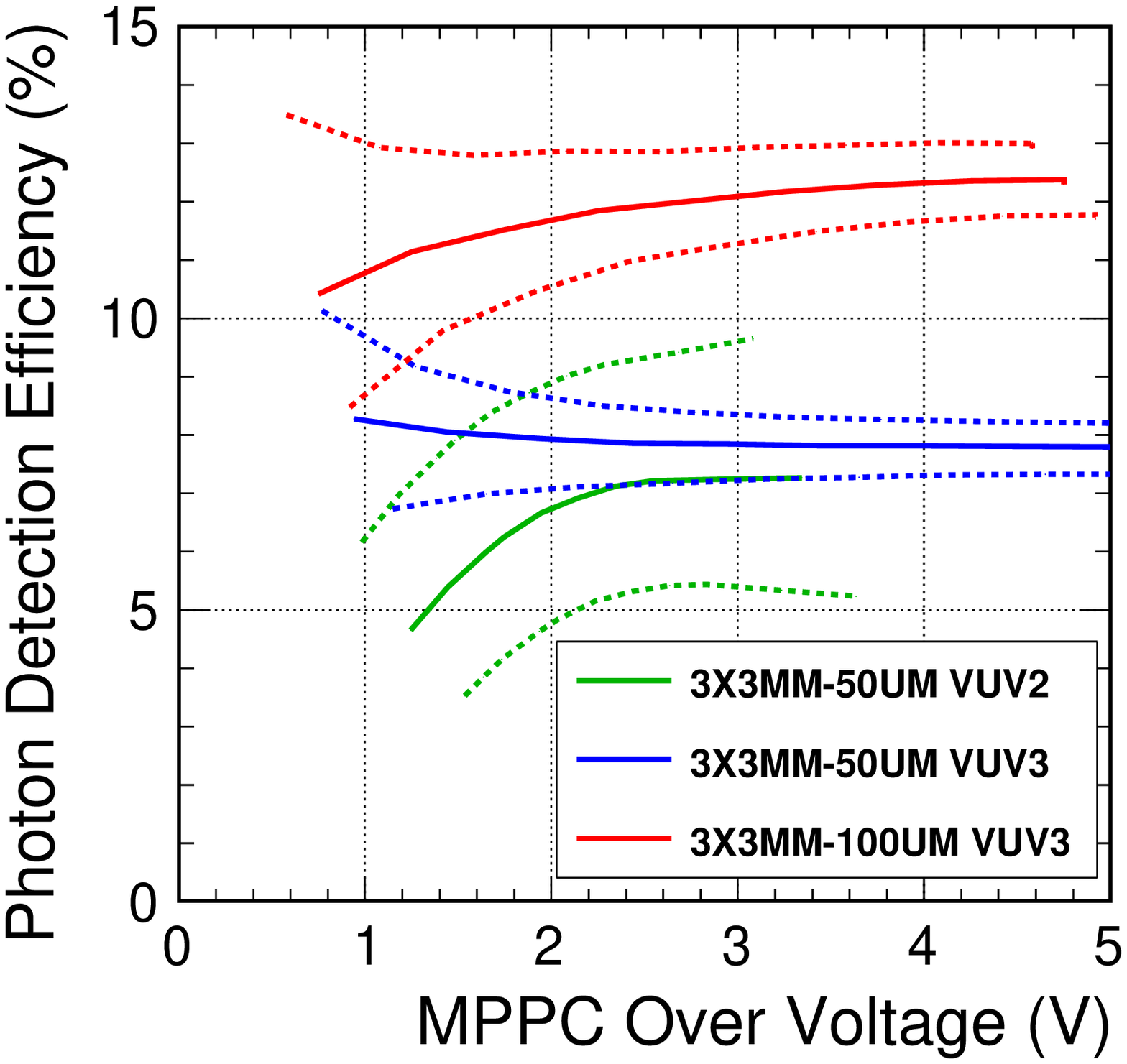}
\end{minipage}
\end{center}
\caption{\label{Liq3}Average number of total pixel hits per 1 photo-electron equivalent signal,
 $N_{\rm pix}$ (left) and corrected PDE (right) as a function of MPPC
over voltage for 128 nm LAr scintillation light.}
\end{figure}

$Q_{\alpha}$ can be translated to the number of photo-electrons by dividing by the gain ($G$). Since the charge distribution
includes the effect of cross-talk and afterpulse, the following correction is necessary to evaluate the number
of real photo-electron of the signal,
\begin{equation}
N_{\rm real} = N_{\rm uncorr}\times\frac{1}{N_{\rm pix}} = \frac{Q_{\alpha}}{Q_{G}} \times\frac{1}{N_{\rm pix}} , 
\end{equation}
where $N_{\rm pix}$ is the correction factor defined as the total number of MPPC pixel hits including
real 1 photo-electron equivalent signal and hits due to cross-talk and afterpulse,
\begin{equation}
N_{\rm pix} = \frac{Q_{\rm LED}}{Q_{G}}\times\frac{1}{\mu}.
\end{equation}
By definition, $N_{\rm pix}$ is always greater than 1 by the effect of the cross-talk and
its over voltage dependence is shown in left plot in Fig.~\ref{Liq3}.
The plot is fitted into a function of sum of constant and exponential, and used for PDE correction.
For example, in the case of V2-50UM-(2), the number of photo-electrons at $V_{\rm ov}$ = 2.3 V 
with and without cross-talk correction are calculated to be $N_{\rm real}=96.7\pm1.2$ and 
$N_{\rm uncorr}=160.5\pm2.0$, respectively.

The expected number of LAr scintillation light at the surface of the MPPC ($N_{\alpha}$) is
estimated using following equation,
\begin{equation}
N_{\alpha} = E_{\alpha}/W_{\alpha}\times A_{\rm MPPC}=5.5~\rm{MeV}
/ (27.5~\rm{eV/photon}) \times 0.7\% = 1400~\rm{photons} ,
\end{equation}
where $E_{\rm \alpha}$ is energy of $^{241}$Am $\alpha$-ray, $W_{\alpha}$ is LAr scintillation
photon emission yield for $\alpha$ particle~\cite{Doke:1999ku}, and $A_{\rm MPPC}$ is acceptance
calculated by solid angle from the $^{241}$Am source to the MPPC.
Finally the photon detection efficiency is determined to be ${\rm PDE}=N_{\rm real}/N_{\alpha}$.
The same procedure is repeatedly performed at different over voltage.
Right plot in Fig.~\ref{Liq3} shows obtained PDEs for three different types of MPPCs as a function of the over voltage.
The PDE is $\sim$8\% for 50 $\mu$m pixel MPPCs, and $\sim$12\% for a 100 $\mu$m pixel MPPC at $V_{\rm ov}=3$ V. 

The basic properties of MPPC in Fig.~\ref{Prop1} (gain and cross-talk) are measured as a function of the over voltage
at the liquid nitrogen temperature, while the $\alpha$-ray data is obtained inside the liquid argon. 
Since the the breakdown voltage shows relatively strong dependence on the temperature ($\sim$50 mV/$^{\circ}$C),
the breakdown voltage at the liquid argon temperature needs to be determined independently.  
Apparently the precision of the breakdown voltage determination ($^{+0.17}_{-0.21}$ V) 
is one of the major source of the systematic uncertainty for the PDE measurement.
Moreover for the V2-50UM type, MPPCs used to observe the LAr scintillation (V2-50UM-(2)) 
and to estimate the gain and cross-talk (V2-50UM-(1)) are different individuals. 
This causes another major systematic uncertainty.
Relative uncertainty on the PDE is assigned to be about 30\% for VUV2 type, and 10$\sim$20\% for VUV3 type 
shown as dotted line in the plot.




\section{Summary}
A new type of the Multi-Pixel Photon Counter (MPPC) which has sensitivity for VUV light
(wavelength $<$ 150 nm) is recently produced by Hamamatsu Photonics K.K..
We have tested the new MPPC under cryogenic temperature ($-$190 $^\circ$C) and measured
several basic properties. Then we have successfully detected the scintillation light of LAr
(wavelength = 128 nm) with this MPPC, and photon-detection efficiency is measured to be $\sim$8\%
for 50 $\mu$m pixel MPPCs, and $\sim$12\% for a 100 $\mu$m pixel MPPC at $V_{\rm ov}=$ 3 V.

For the recent WIMP dark matter search experiments using liquefied noble gas (argon and xenon),
it is reported that the background events at the surface of the detector may be mis-reconstructed
as events at center of the detector. Those mis-reconstructed events remain in the signal
region and limit the physics sensitivity.
For the ANKOK experiment, we are considering to improve the spatial reconstruction resolution
by arranging the MPPC. For the double phase argon detector the spatial resolution in terms of
electron drift direction is determined very precisely ($O$($<$mm)) by the time difference between
direct scintillation and secondary scintillation caused by the drifted electrons (time projection).
However the resolution in transverse plane is limited by the size of PMT (typically 3 inches).
Locating the small size of MPPC (3 mm$\times$3 mm) at the detector wall in gas argon phase would
potentially improve the spatial resolution to distinguish the background events from the wall
since the VUV MPPCs are expected to detect direct 128 nm scintillation light that contains
more emission position information than wavelength shifted light by TPB.
To conclude, this kind of new MPPC may be considered to open new style of argon scintillation
detector in the near future. Thus further development toward practical stage including low-background
technique is desired.

\section*{Acknowledgments}
We first acknowledge the solid state division of Hamamatsu Photonics K.K. for providing us with
the new VUV MPPC samples. We particularly appreciate useful discussions with Y.~Hakamata, K.~Sato,
and R.~Yamada. We are grateful to W.~Ootani for helpful comments and support.
This work is a part of the outcome of research performed under a
Waseda University Research Institute for Science and Engineering (Project numbers 13C09 and 14C12),
supported by JSPS Grant-in-Aid for Challenging Exploratory Research Grant Number 25610060.

\smallskip

\end{document}